\documentclass[fleqn,10pt]{wlscirep}
\usepackage{multirow}
\usepackage{bm}
\usepackage{subcaption}

\title{A Multilayer perspective for the analysis of urban transportation systems}

\author[1,*]{Alberto Aleta}
\author[1,2]{Sandro Meloni}
\author[1,2,3,$\dagger$]{Yamir Moreno}

\affil[1]{Department of Theoretical Physics,
Universidad de Zaragoza, Zaragoza 50009, Spain}
\affil[2]{Institute for Biocomputation and Physics of Complex Systems (BIFI), Universidad de Zaragoza, 50018 Zaragoza, Spain}
\affil[3]{Institute for Scientific Interchange, ISI Foundation, Turin, Italy}

\affil[*]{albertoaleta@gmail.com}
\affil[$\dagger$]{yamir.moreno@gmail.com}


\begin{abstract}
Public urban mobility systems are composed by several transportation modes connected together. Most studies in urban mobility and planning often ignore the multi-layer nature of transportation systems considering only aggregated versions of this complex scenario. In this work we present a model for the representation of the transportation system of an entire city as a multiplex network. Using two different perspectives, one in which each line is a layer and one in which lines of the same transportation mode are grouped together, we study the interconnected structure of $9$ different cities in Europe raging from small towns to mega-cities like London and Berlin highlighting their vulnerabilities and possible improvements. Finally, for the city of Zaragoza in Spain, we also consider data about service schedule and waiting times, which allow us to create a simple yet realistic model for urban mobility able to reproduce real-world facts and to test for network improvements.    
\end{abstract}
\begin{document}

\flushbottom
\maketitle
\thispagestyle{empty}

\section*{Introduction}

Multiplex networks\cite{KivelaeArenasBarthelemyEtAl2014} are useful representations of systems in which the same set of nodes may be connected by different types of relationships.  Examples of systems that can be modeled as multiplex networks include social networks, transportation systems with multiple transportation modes or biological systems in which different types of interactions are accounted for\cite{KivelaeArenasBarthelemyEtAl2014}. In a multiplex network, nodes and links are grouped in layers according to their nature. Layers can be interdependent and they contain information which would be lost if we only considered the corresponding aggregated network. It has also been shown recently that different types of dynamics that are run on top of multilayer systems also provide new insights into the problems being modeled \cite{BuldyrevParshaniPaulEtAl2010,CozzoPRE2012,CozzoPRE2013}.

As discussed in previous papers\cite{Gallotti2014} most available studies on urban transportation consider either one single transportation mode or many modes but all merged in one aggregated network. Thus, the introduction of the new framework of multiplex networks for the analysis of urban transportation systems might allow to better understand complex issues like how to accurately account for the interplay between different transport modes. However, even though few works started to use a multiplex representation to study  failures\cite{DeDomenico2014} and efficiency\cite{Gallotti2014,Gallotti2016} in transportation systems, they still represent isolated cases. For instance, there are very recent studies that rely on a complex notation to incorporate multiple modes\cite{Zhang2014}, but that simply aggregates the whole network, thus losing information regarding transfer times\cite{Pang2015}. Very up-to-date reviews where the term multilayer is either not present or used in a completely different way\cite{Lin2013,Ibarra-Rojas2015} can also be found in the specialized literature.

The few previous studies on urban transportation systems as multiplex networks focus on addressing their multimodal nature, considering each layer as a transportation mode, to study their resilience\cite{DeDomenico2014} or their coupling\cite{Strano2015}. In this way, all the lines of the same mode (i.e. buses, metro and tram) are aggregated in a sort of {\em superlayer}.  This representation is extremely compact --with only few layers-- but it totally neglects transfer and waiting times between lines of the same mode that could eventually lead to a wrong estimation of shortest paths or travel times. Another solution is to consider each line of each mode as a single layer. While in this case we can preserve transfer times and synchronization between stops on different lines,  it is not possible to quantify the importance of a transportation mode for the mobility in the system. 

To reconcile both approaches, in this paper we propose an urban transportation model based on multiplex networks where both representations are used to extract different information from the system. We show that {\em superlayers} are fundamental to study the interdependency and resilience of the system while to have a realistic modeling of human mobility the single line per layer perspective should be adopted. We test our model using $9$ different urban transportation networks raging from small cities of few hundred thousand inhabitants to megacities like London or Berlin. Finally, for a medium size city we will introduce detailed data about schedule and transfer times to create a more realistic dynamical scenario to test against real-world experimental data and facts. The remaining of the text is organized as follows. First, we give an overview of our multiplex representation in the Methods section and then use it to study the structure of $9$ urban transportation networks in the first subsection of results. Finally, in the second part of the Results section, we focus on one case study (city) for a deeper analysis. Specifically, we test if the model is able to reproduce experimental data and check for its possibilities regarding the study of service disruptions and network improvements.

\section*{Methods}

We start considering each line of each mode of transport as a single layer. Each stop will be a node and there will be a weighted link between two nodes on a layer if the corresponding line passes through both of them, being the corresponding geographical distance between them the weight. Although the same bus stop may be present in multiple layers, allowing transfers between them, this might not be the case between layers of different modes. To solve this problem, we will connect (with inter-layer links) each node of one mode to the closest one of each of the other modes as long as the distance between them is less than $100$ m. 

In the second part of the results section, however, we will follow a slightly different scheme to add more real features to the model. In this case, we will add a new layer which represents the land to introduce the possibility of moving through the city by walking. To create this layer we first took the population density grid of the European Union \cite{Gallego2010}, which is composed of roughly $100 \times 100$ m cells, and extracted those inside the area of interest. Then, we took those cells which had a population density greater than zero and set a node in the middle of them. Finally, we linked together  nodes belonging to neighboring cells and added the corresponding weight  (distance in meters). To connect this layer with the rest of the system we simply determined which cell each stop belongs to and establish a link between that stop and the corresponding node of the land layer. Once this was done we added, again, the distance between stops as a weight to their links. This distance is simply the geographical distance in the case of tram and metro, but for bus stops it was calculated taking into account street patterns using the Google Maps API\cite{GoogleAPI}. In this way, the second model is much more powerful, allowing us not only to check the validity of the conclusions extracted from the first analyses in a more realistic scenario but also to study more complex phenomena as service disruptions.

\section*{Results}

\subsection*{Structure of urban transportation networks}

In table \ref{Tabla} we present the principal characteristics of the networks we are going to analyze. Data were obtained from a variety of sources, from each company's website to city's open data portals, and then arranged as described in the model discussed in Methods. As we can see they are very different in size and composition. For example, while Vitoria has a population of roughly $250,000$ individuals and a small transportation network consisting of $302$ nodes and $16$ layers, London is one of the biggest cities in Europe with more than $8\cdot 10^6$ inhabitants and a network consisting of $19,459$ nodes and $555$ layers.

\begin{figure*}
	\begin{subfigure}{.49\textwidth}
		\centering
		\includegraphics[width=\linewidth]{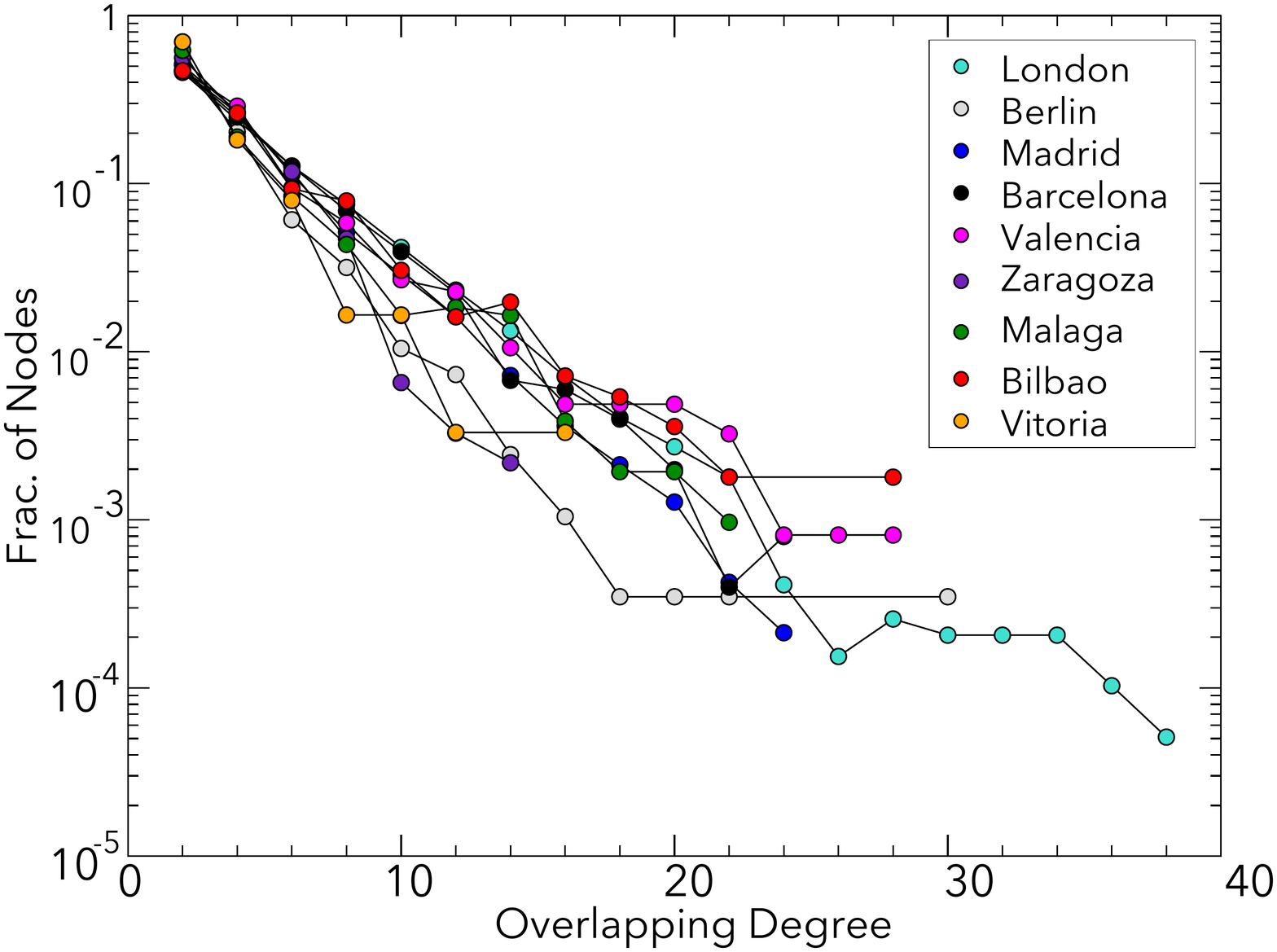}
	\end{subfigure}
	\begin{subfigure}{.49\textwidth}
		\centering
		\includegraphics[width=\linewidth]{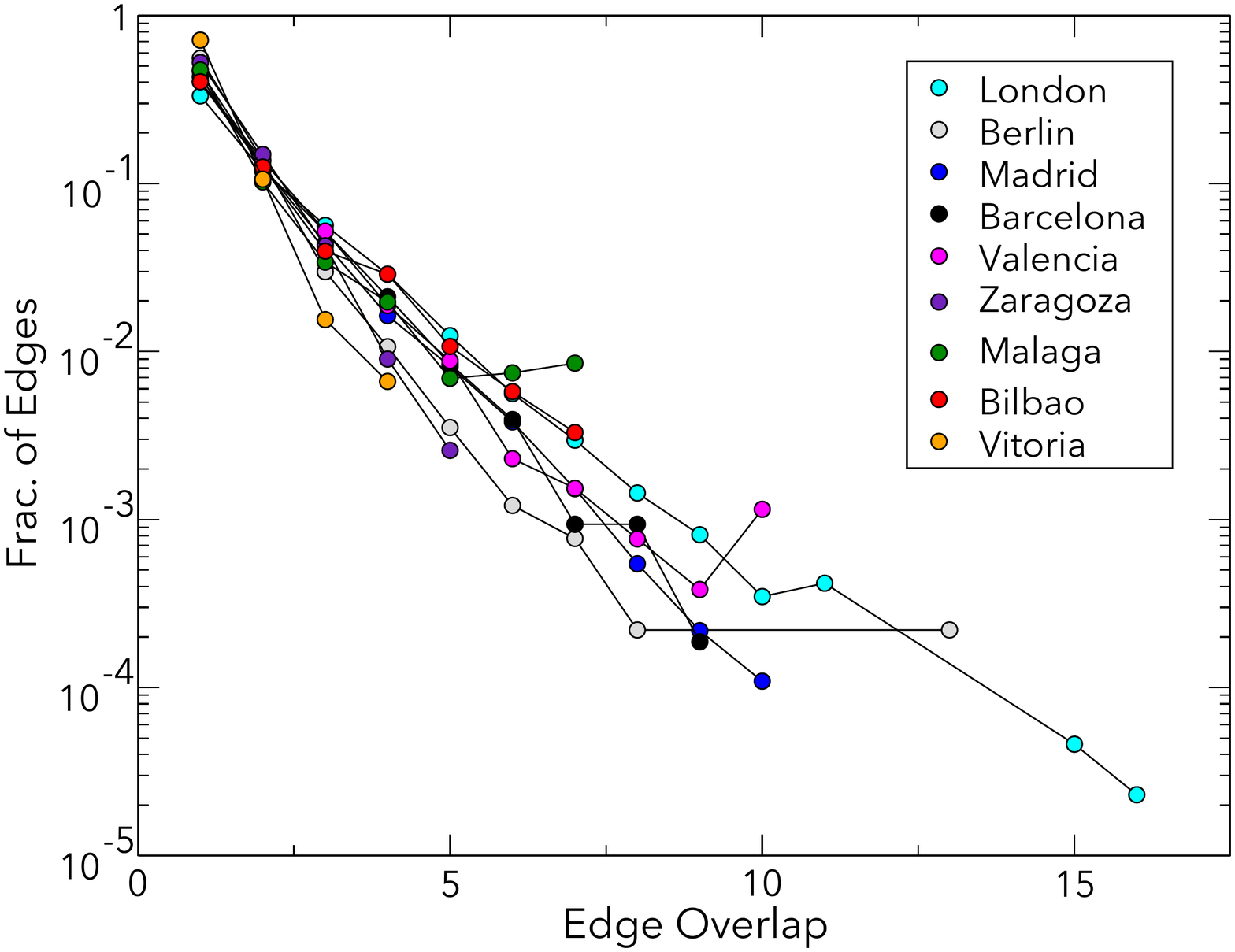}
	\end{subfigure}
    \caption{Left: overlapping degree distribution of each network. Right: edge overlap distribution of each network.  Despite these networks being quite different in composition and size they share some universal properties.}\label{fig:overlapping}
\end{figure*}

We start our analysis with one of the most basic measures of graph theory, the degree. In multiplex networks it can be defined in multiple ways. A straightforward approach is to consider the degree of node $i$ as a vector $\bm{k}_i$ of length $M$ (the number of layers) where each element $j$ represents the degree of node $i$ in layer $j$.  However, in this particular case this measure does not provide much information because, as each layer represents a single line, if a node is present in one layer it will have degree 2 in that layer (or 1 in special cases as the first/last stop of a line, although in our networks both stops are always the same one). What we can do is to examine the overlapping degree, $o_i$, which is simply the sum of the elements of $\bm{k}_i$ \cite{Battiston2014} (Figure \ref{fig:overlapping}).

As we can see, although cities are very different from each other their overlapping degree distribution is quite similar: most of the nodes are only present in one layer ($o_i=2$), some are present in two layers ($o_i=4$) and only a few can be found in three or more layers. Given the nature of the system one could expect a low level of overlapping between layers. However, what is really interesting is the fact that the maximum overlapping is quite similar in all the networks, even though their number of layers (and thus their theoretical maximum overlapping degree) differs a lot. Similar results are found if we look at the edge overlap distribution, where the edge overlap, $o_{ij}$, is defined as the number of layers where a link between nodes $i$ and $j$ exists \cite{Battiston2014}. This kind of universal behavior can be understood if we take into consideration that these networks are embedded in city space and thus the real theoretical maxima for both the overlapping degree and the edge overlap are not given by the number of layers but by either physical constraints or citizens' interest.

\begin{figure*}
\centering\includegraphics[width=0.7\linewidth]{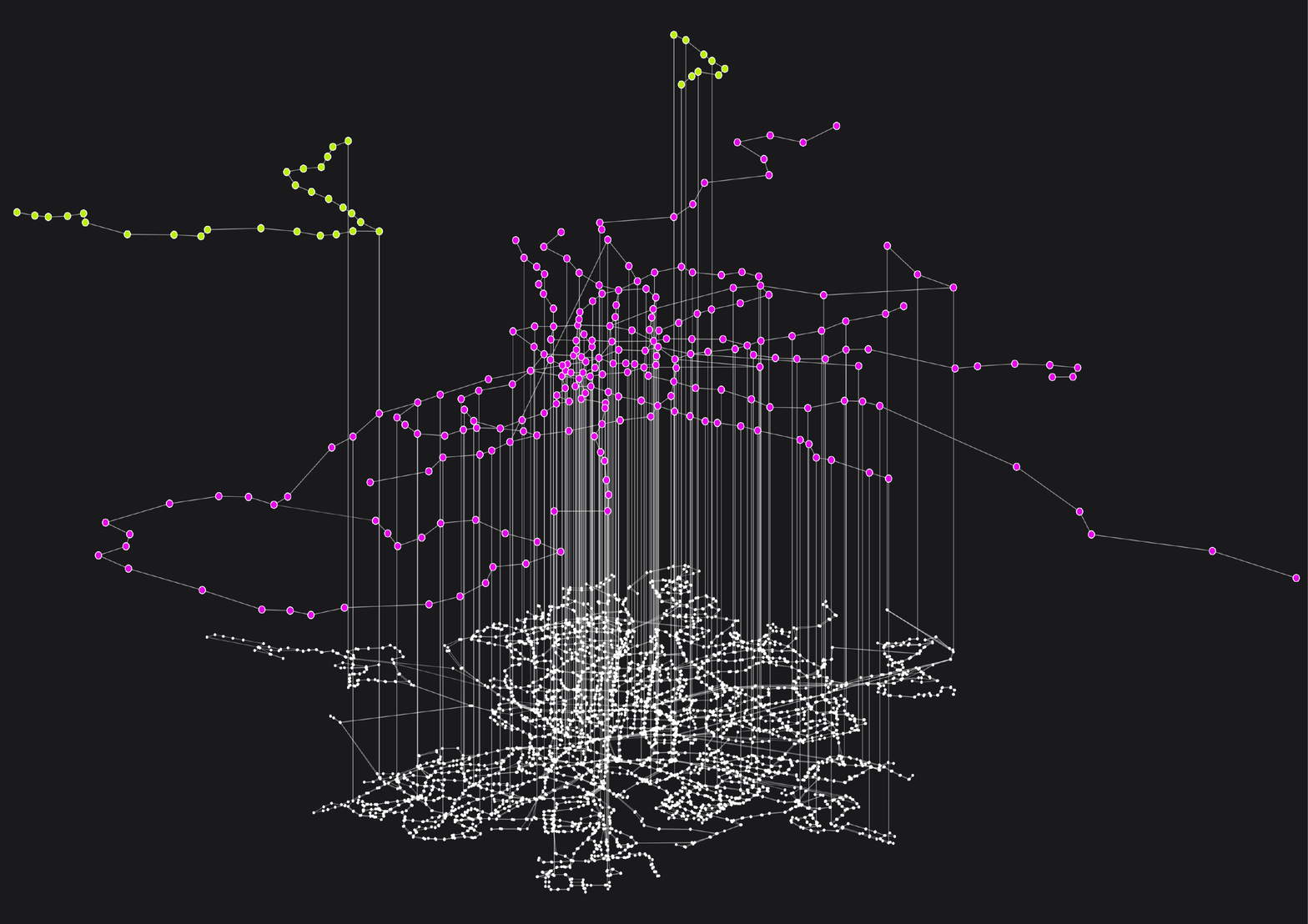}%
\caption{Superlayer representation of the Madrid transportation system. The figure represents the three transportation modes considered: tram (yellow nodes, upper layer), metro (purple nodes, mid layer) and buses (white nodes, bottom layer). See Table\ref{Tabla} for statistics of these layers.}%
\label{superlayers}%
\end{figure*}

An interesting feature of multiplex networks is that the distribution of a quantity across the layers is, at least, as important as the overall value. For example, one node can have high overlapping degree either becau0se it has a low value in all the layers or because it has a high value in just a few layers. However, in our model this does not apply due to the singularities of public transportation networks  that we discussed before. Nevertheless, to get  insights about  the importance of a transportation mode over the others we can  switch perspective and consider the {\em superlayers}  representation, see Fig.\ \ref{superlayers}. As in recent works\cite{DeDomenico2014,Gallotti2014,Gallotti2016}, we propose to group together lines belonging to the same transport mode ending up in most of the cases with three {\em superlayers} representing bus, tram and metro lines respectively. Except for this modification the other elements of the model remain untouched with interlayer links connecting nearby stops of different modes.

\begin{figure*}%
\centering\includegraphics[width=\linewidth]{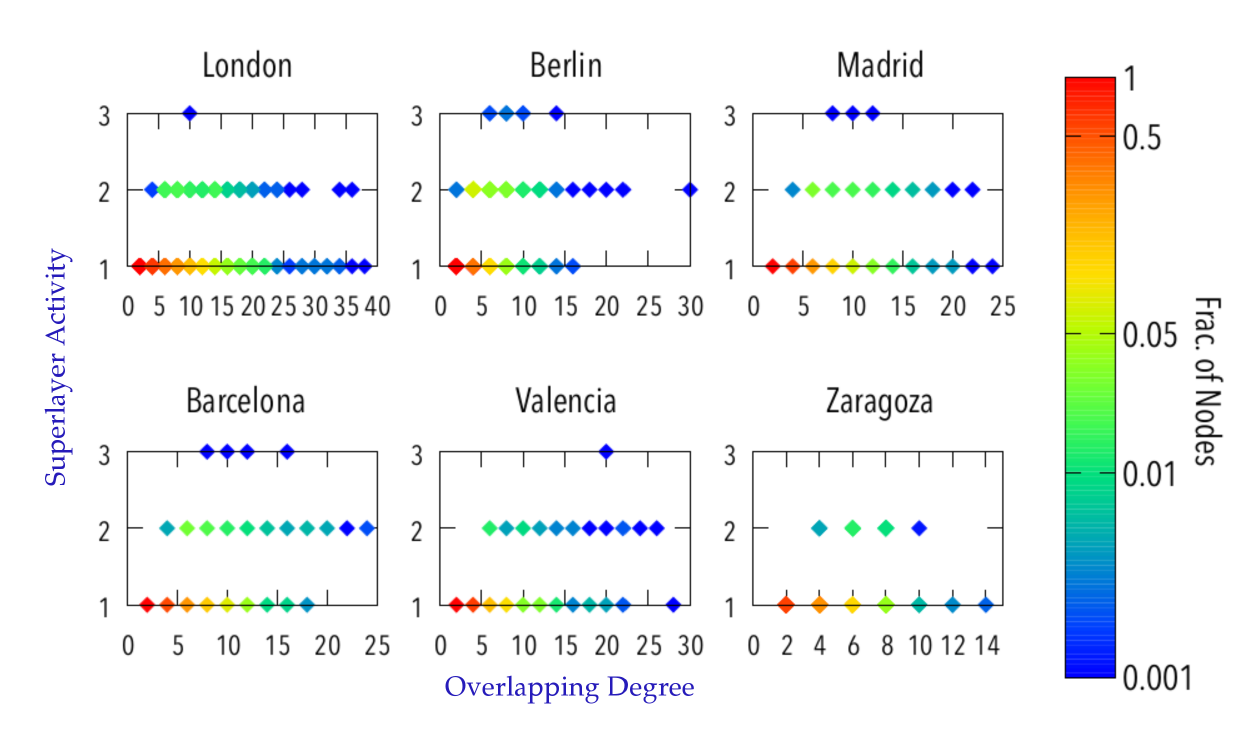}%
\caption{Superlayer activity versus overlapping degree for each network. }%
\label{fig:degreeactivity}%
\end{figure*}

To start this new analysis let us denote by $C_b$, $C_t$ and $C_m$ the subsets of layers corresponding to bus lines, tram lines and metro lines respectively. Now we can redefine the overlapping degree as $o_i = \sum_x o_i^x$ where $o_i^x = \sum_{\alpha \in C_x} k_i^{[\alpha]}$ with $x=\{b,t,m\}$. Then, instead of considering the activity distribution across layers \cite{Nicosia2014}, $B_i = \sum_\alpha (1-\delta_{0,k_i^{[\alpha]}})$, we can study the activity distribution across superlayers, $B'_i = \sum_x (1-\delta_{0,o_i^x})$, which we represent versus the overlapping degree in Figure \ref{fig:degreeactivity}.

At first one may think that nodes with the highest overlapping degree would also be those with the highest superlayer activity, but as we can see this is not the case. In fact, stops belonging to just one superlayer are the ones which tend to have the maximum overlapping degree.  Those nodes  despite being present in only one superlayer, surely have a major role for the mobility of the system. But, on the other hand, if we think in a disruption of the system it will be much easier to, for example, move temporarily a bus stop to a street nearby, even if it has a lot of lines, than cope with a disruption in a metro or tram stop. Thus, there is not a clear answer to the question of which node is the most important in these networks, as it depends on what we consider important. 

To end this structural study we will focus on another measure used in multiplex networks analyses: interdependence \cite{Nicosia2013}. Interdependence of node $i$, $\lambda_i$, is defined as the sum over every other node $j$ of the fraction of shortest paths between node $i$ and $j$ which go through two or more layers over the total number of shortest paths between them. Hence, if $\lambda_i$ is close to 0 it means that most of the paths go through just one layer while if it is close to 1 most of the paths go through 2 or more layers. The network interdependence is obtained as the average over all nodes. However, even though this measure is quite interesting as it provides some information which can not be obtained using the aggregated network alone, in our system it is not necessary because we already know that layers are interdependent. Indeed, if we look at table \ref{Tabla} we can see that most of the networks have a number of nodes of the order of 1000, but a single bus line usually has around 50 stops, 100 at most, and lines of tram and metro have even less. This means that to reach every node of the network we will surely have to cross multiple layers regardless of our starting position. One possible solution would be to work with superlayers instead of single layers, as they would be denser. But then we would face a similar problem as the bus superlayer is much bigger than the other two.

Therefore, we slightly modify this metric to take into account the specific nature of transportation networks. Let us denote by $\psi_{ij}(\alpha)$ the number of shortest paths between nodes $i$ and $j$ which go across two or more layers being $\alpha$ one of them and by $\sigma_{ij}$ the total number of shortest paths between $i$ and $j$. We can define the interdependency of layer $\alpha$ as:
\begin{equation}\label{eq:interdep_layer} \lambda'_\alpha = \frac{\sum_i \sum_{i\neq j} \psi_{ij}(\alpha)}{\sum_i \sum_{i\neq j} \sigma_{ij}}, \end{equation}

\noindent
that gives us the fraction of all the shortest paths of the system which go through layer $\alpha$. Although this can give us a lot of information if we want to study problems like congestion in a particular system, for this general analysis we can go one step further and define the interdependency of superlayer $x$ as:
\begin{equation}\label{eq:interdep_superlayer}\lambda'_x = \frac{\sum_i \sum_{i\neq j} \psi_{ij}(x)}{\sum_i \sum_{i\neq j} \psi_{ij}},\end{equation}

\noindent
that tells us how many shortest paths in the system have, at least, one link in superlayer $x$. Note that in this case, we normalize only over those shortest paths which use two or more layers because we are interested in figuring out if, given the need to change to another line, the tendency is to remain in the same transport mode or the system tends to be multimodal. 

\begin{figure*}
\centering\includegraphics[width=\linewidth]{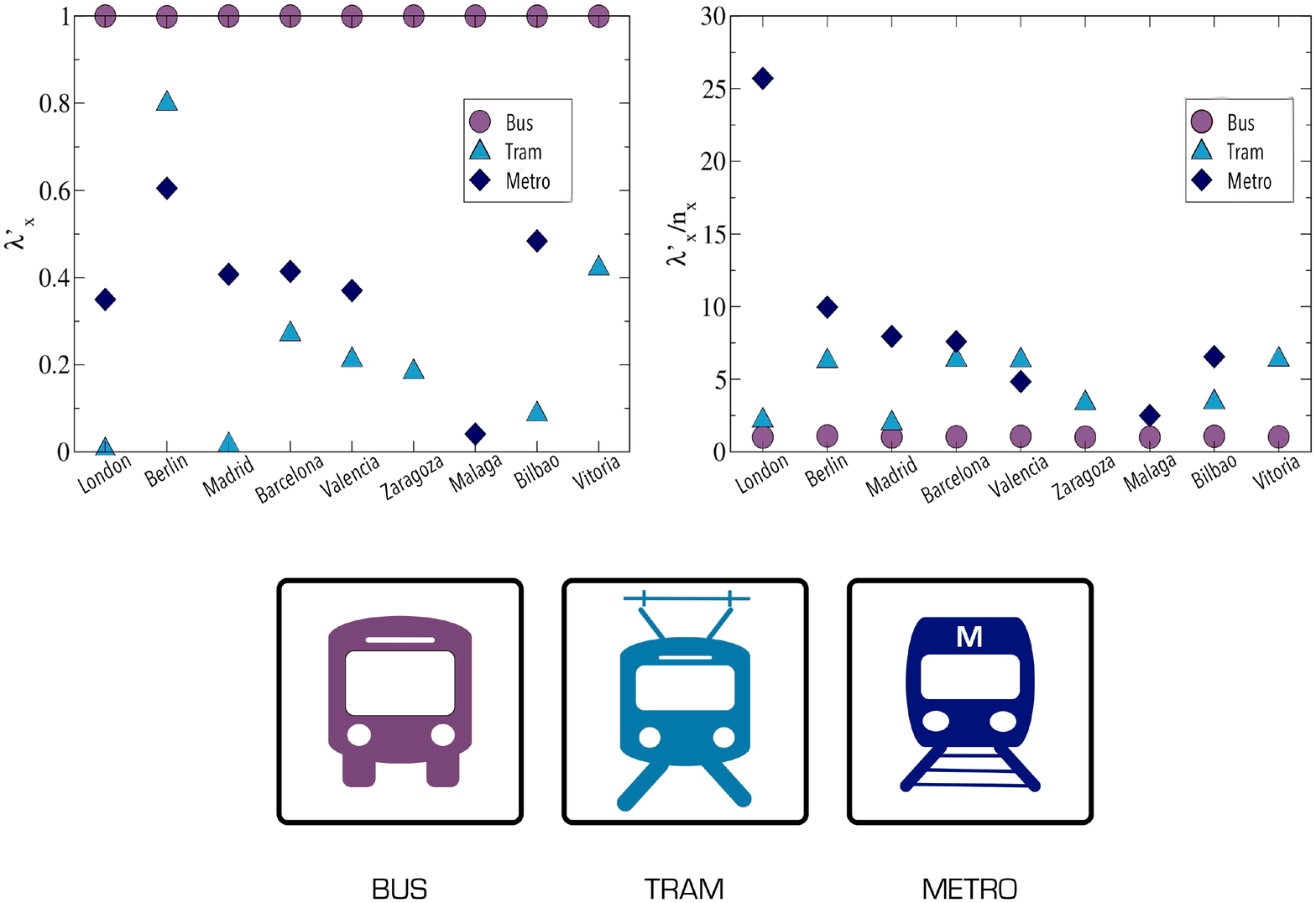}%
\caption{Left: Superlayer interpendency as defined in \eqref{eq:interdep_superlayer}. Right: Superlayer interdependency divided by $n_x$ which is the sum of the weights of all the links inside superlayer $x$ over the sum of the weights of all the links in the system.}%
\label{fig:interdep}%
\end{figure*}

Results of this measure are shown in Figure~\ref{fig:interdep}a. Firstly, we note that almost all the shortest paths under consideration have at least one link in the bus superlayer, but this is quite logical given that the bus superlayer is much bigger than the others. Thus, most of the shortest paths will start or end in this superlayer. However, a closer look reveals an interesting result. Take the case of Madrid, for example. Even though its metro superlayer has only 16 layers and 241 nodes, while its bus superlayer has 177 layers and 4590 nodes, more than 70\% of the shortest paths have at least one link in it. Similar results are found in the rest of networks as, for example, in Zaragoza where 20\% of the shortest paths make use of the tram which has only 1 line and 50 nodes, with its bus superlayer being composed by 35 layers and 902 nodes.

As it can be seen, to fully understand these results it is necessary to take into account the size of the superlayers, the problem is how to define it. In this case, as we are exploring paths from node to node, we will consider the fraction of all the nodes in each superlayer as a measure of size. This way, if $n_x$ is the fraction of nodes in superlayer $x$, we will divide $\lambda_x$ by $n_x$ to obtain the desired result. This procedure has it drawbacks as it is not upper bounded, but on the other hand it allows us to extract information on the importance of each layer in an easier way.

From this modified measure (see Fig.~\ref{fig:interdep}b), we observe that the tram and the metro modes are of utmost importance for the mobility of these systems, as they are part of much more shortest paths than it would correspond judging by their size. Note that we have not taken into account that they may have, on average, higher speed or greater carrying capacity than buses, and hence, the previous results are obtained only considering a topological point of view. The reason seems clear: metro and tram are usually used to connect distant points with straighter routes than bus lines. Another interesting conclusion that is a consequence of the previous one is that, although most of the networks examined rely mainly on two transportation modes, urban transportation is quite multimodal. A good example of this is Madrid's network where bus nodes cover most of the city and metro nodes connect distant locations with straighter paths throughout it. However, only one of the three tram lines overlaps with bus nodes, the others seem to go to locations that are not covered by bus. Thus, the tram mode is not used as a way to reach certain locations faster but to just connect distant locations at the periphery of the city.

\subsection*{Case study: realistic modeling of an urban transportation system}
 
In this section, we study a detailed model for urban transportation that includes not only the structure of the networks but also data about frequencies and traveling speeds. Our aim here is to realistically mimic a scenario that allows studying the efficiency of an urban transportation system and its response to malfunctioning or improvements. To do so, we also include the land layer, as discussed in Methods, allowing passengers to get to the stop which might be across the street even if there is not a direct link between different lines. In this way, we will be able to simulate paths starting/ending at any point of the city and not only at lines stops.

We take the transportation network of the city of Zaragoza (Saragossa) in Spain as our case study, since we have data regarding average speed and frequency for each line\cite{velocidad,frecuencia}. This allows us to consider time, so that now the shortest path will be the one which takes less time, instead of the geographical distance. To this end, we divide the weight of the links by the average speed on their layer. Thus, the weight of a link will now represent time. Moreover, all weights are fixed throughout the simulation, as we consider that the speed is always the same, except for links connecting land nodes and stops. In the latter case, the weight of these ones will be the time at which the next vehicle will arrive minus the current time: therefore, given a certain path, the sum of the weight of the links used that belong to the land layer, any other layer or to the inter-layer links set will give the total time spent walking, in a vehicle or waiting, respectively.

Even more, due to the recent construction, in $2011$, of the tramway, surveys were carried out regarding the impact of this new transportation mode in the mobility of the city\cite{survey}. Thus, we are able to test whether our model can provide similar (qualitative) results to passengers' experience, namely: (i) mobility hinges on the tram, (ii) a lot of transfers are needed, (iii) if there is a disruption in the tram line the whole system is affected, (iv) bus stops are far from their destination and (v) bus frequencies are low. At the same time, although approximately, the previous effects can be quantified. 

As we do not have data regarding passengers flux we do not take into account the carrying capacity of the vehicles and thus we  consider a free flow regime as well as two different scenarios: movement from any point of the city to the city center (coordinates 41.652, -0.881) and vice versa and from any two points located at least 2 km away from each other, which has been reported as the minimum distance a passenger has to go to consider using public transportation\cite{Distancia}. 

\begin{figure*}
\centering\includegraphics[width=\linewidth]{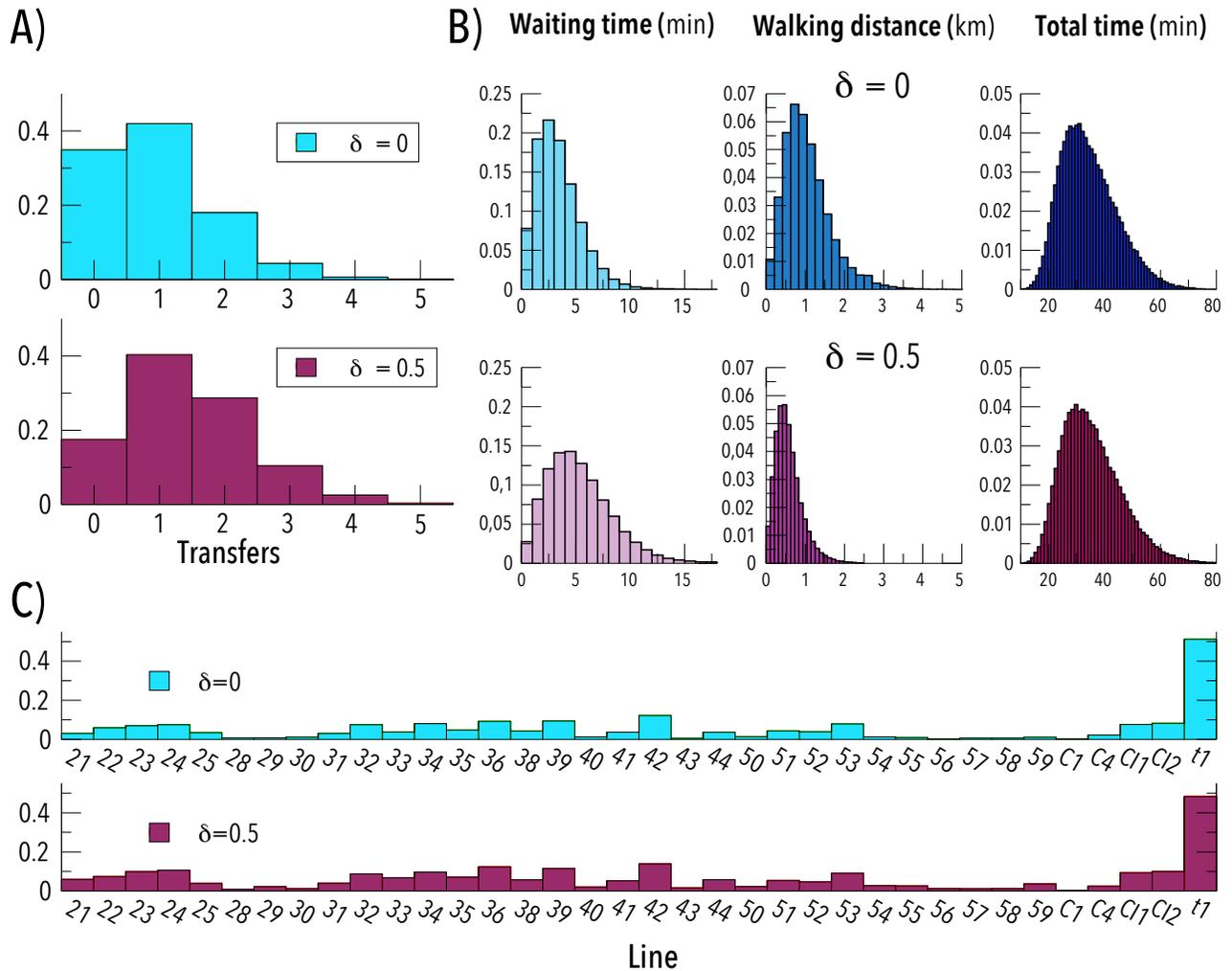}%
\caption{Overview of the mobility of the system with random origins and destinations: (a) number of transfers made by individuals to reach their destination; (b) the time each individual has been waiting in any stop (left), the distance covered by walking (center) and the total time of the trip (right); (c) fraction of individuals who have used each line. Note that it is normalized over the number of travels thus, as a passenger may use more than one line, the total sum is not 1. The parameter $\delta$ denotes a penalization in the walking speed in order to force the use of transports.}%
\label{fig:big}%
\end{figure*}

In Figure~\ref{fig:big} we show an overview of the results obtained considering 1.000 individuals per minute with random origin and destination between 08:30 am and 10:00 am with a walking speed $5$ km/h\cite{Browning2006}. In Fig.~\ref{fig:big}(a) the distribution of the number of transfers is shown, note that only 35\% of the individuals have reached their destination without transfers, which agrees with (ii). In Fig.~\ref{fig:big}(b) we represent the distributions of the waiting times, total time and distance covered by walking. With a walking speed of $5$ km/h, 1 km is equivalent to 12 minutes and thus it represents approximately one third of the total time, which agrees with (iv) but not with (v). Now, suppose that an individual needs to do one transfer to get to his location. If one of the two frequencies is too low, it might be faster to walk to get to the second line instead of getting there using another line or equivalently walking from the transfer point to his destination instead of waiting for the second vehicle. He would see two problems, he needs to walk a lot and the frequencies are low, but as he has not been waiting it will not be reflected on our model.

To avoid this problem and following other studies on urban mobility\cite{Browning2006} we introduced another parameter $\delta$ to module the walking speed $v_w$ such as $v_w=(1-\delta)\cdot 5$ km/h. By tuning its value we can prevent  individuals to avoid transfers but, at the same time, we still allow them to do so if the distance is small enough. As we can see in Fig.~\ref{fig:big}(a) with $\delta=0.5$ the number of transfers increases greatly, which means that large walking distances are not present anymore. If we look at time we see that, as expected, waiting time increases and the distance covered by walking reduces. The total time has been readjusted taking into account the decrease in walking speed so that if it varies is only because of the new transfer and, surprisingly, it almost does not change. This means that we have a duality in the system as we can choose between walking or waiting while keeping the total time constant, which indeed agrees with (iv) and (v). Note also that,  although the waiting time distributions may not seem quite high if we compare them to the total time distributions, several studies have shown that waiting time perception is usually overestimated, specially if the real time has been quite low \cite{Mishalani2006,WatkinsFerrisBorningEtAl2011,Lagune-Reutler2015}. Finally, picture Fig.~\ref{fig:big}(c) shows the fraction of individuals which make use of each line normalized over the total number of trips. In both scenarios almost half of the trips use the tram line which completely agrees with (i). Even more, it also agrees with what we saw when we studied the structure of the network in the previous subsection. 

\begin{figure*}
\centering\includegraphics[width=0.75\linewidth]{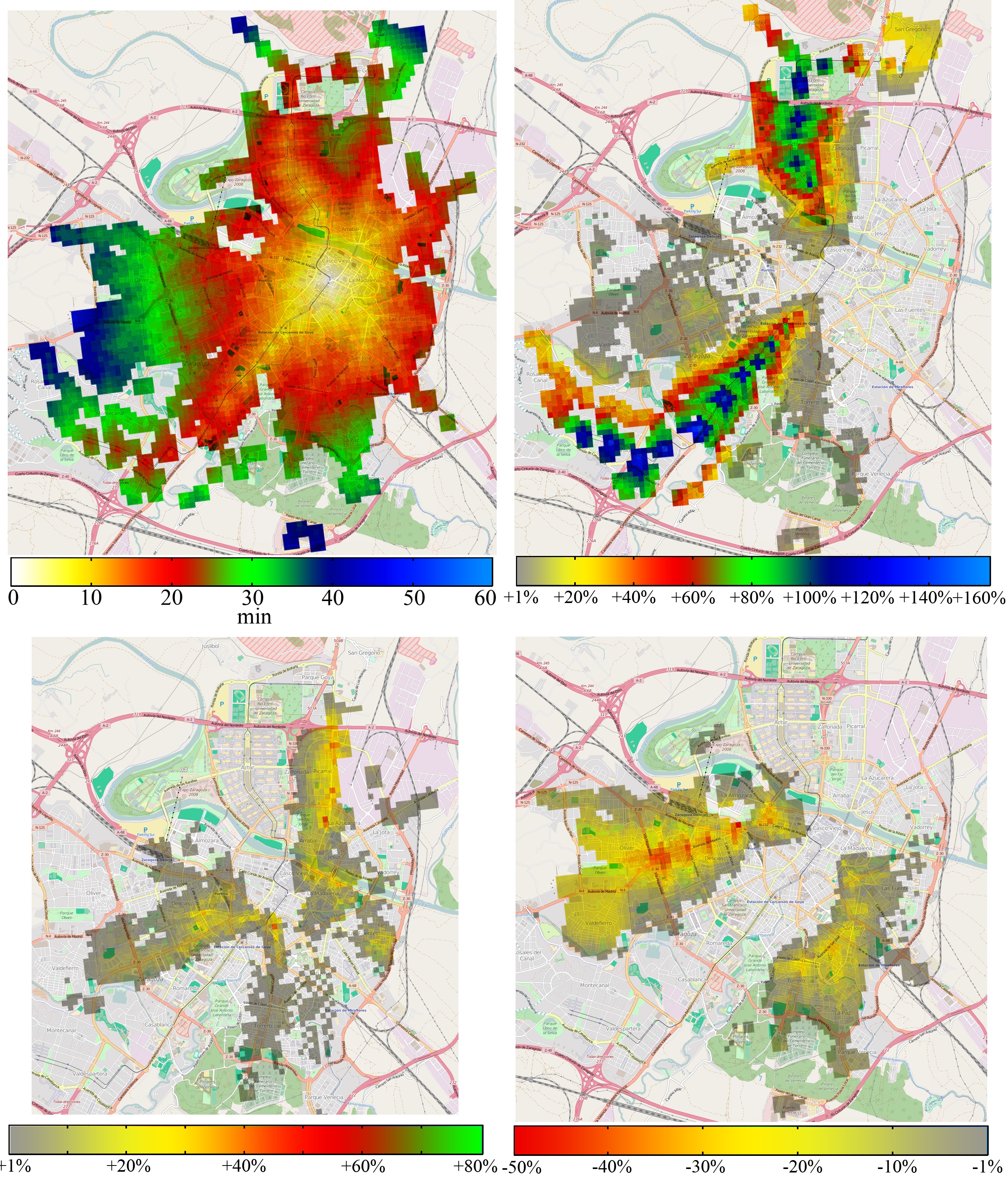}%
\caption{Top left: average time it takes to get to the city center between 08:30 and 09:00. Top right: difference with the original case when the tram is removed. Bottom left: difference when the 2 most used lines to get to the city center are removed (lines 22 and 35). Bottom right: difference if we add a new tram line. This pictures have being done using tiles from OpenStreetMap \cite{OSM}.}%
\label{fig:maps}%
\end{figure*}

The last item left to check is (iii). For simplicity, we will focus only on trips to the city center and vice versa (Figure~\ref{fig:maps}). On the top left panel we show the average time it takes to get to the city center between 08:30 and 09:00 am.  Now, our model allows us to easily test the behavior of the system during service disruptions by just removing the affected nodes or layers. In the top right panel of Fig.~\ref{fig:maps} we show the increase on the total time if we remove the tram line completely, result that agrees with (iii). Although the tram line is quite vulnerable as a disruption on a single node may cause the whole line to fail (as it follows a fixed path), it is important to note that this line has a couple of links which allows it to run in loops and thus, in this particular scenario, only the northern or the southern parts would be affected and not the whole line. Nevertheless, if we look at the bottom left panel of Fig.~\ref{fig:maps} we see that the situation is completely different if we remove the two most used bus lines to get to the city center (lines 22 and 35). As the bus mode is more redundant, the effect is much lower, at least under free flow conditions. Besides, a disruption on a single bus node does not cause so many problems on the whole line as it can be easily moved to a location nearby. Thus, we can say that the bus network is much more resilient while tram lines speed up the complete network.

To conclude,  our model can also be used to easily test network improvements as the addition of new lines. On the bottom right panel of Fig.~\ref{fig:maps} we show the differences on the average time to get to the city center if we add a new tram line from east to west as it has been recently proposed by the city council\cite{linea2}. As we can see the west part of the city would be the most benefited by this addition as the decrease in time is higher than in the rest of the city, even more taking into account that, as shown in the top left panel of Fig.~\ref{fig:maps}, that part of the city was further away from the city center. Note that we have not removed any bus line and thus this result shows us again how tram and metro lines naturally speed up the network.

\section*{Discussion}

In this paper, we have proposed a method to model public transportation systems as multiplex networks, which allows to either get more insights into their network properties or extract new conclusions of practical value. We have analyzed the structure of $9$ urban transportation networks and found universal properties that can be related to the underlying structure of the cities. We have also shown that both a {\em per line} and a {\em per transportation mode} representations are useful and complementary to extract information about the functioning of transportation systems and to assess their vulnerabilities. Finally, using detailed data about service schedule and waiting times we created a realistic model for urban mobility. Despite its relative simplicity, we showed that our model not only reproduce real world facts, but that it can also be used to explore important issues like the impact of service disruptions and ways for network improvements, using information which, maybe with the exception of the average speed of each line, should be publicly available for most major cities. Needless to say, a deeper analysis would require to include the street network as the land layer and some information that might be harder to find -such as traffic light or mobility patterns-, but that is beyond the scope of this study. Concluding, the proposed model can be used to for a first diagnosis of the state of any urban transportation network using publicly available information and few computational resources.

\bibliography{referencias}

\begin{thebibliography}{10}
\expandafter\ifx\csname url\endcsname\relax
  \def\url#1{\texttt{#1}}\fi
\expandafter\ifx\csname urlprefix\endcsname\relax\def\urlprefix{URL }\fi
\providecommand{\bibinfo}[2]{#2}
\providecommand{\eprint}[2][]{\url{#2}}

\bibitem{KivelaeArenasBarthelemyEtAl2014}
\bibinfo{author}{Kivel{\"a}, M.} \emph{et~al.}
\newblock \bibinfo{title}{Multilayer networks}.
\newblock \emph{\bibinfo{journal}{Journal of Complex Networks}}
  \textbf{\bibinfo{volume}{2}}, \bibinfo{pages}{203--271}
  (\bibinfo{year}{2014}).

\bibitem{BuldyrevParshaniPaulEtAl2010}
\bibinfo{author}{Buldyrev, S.~V.}, \bibinfo{author}{Parshani, R.},
  \bibinfo{author}{Paul, G.}, \bibinfo{author}{Stanley, H.~E.} \&
  \bibinfo{author}{Havlin, S.}
\newblock \bibinfo{title}{Catastrophic cascade of failures in interdependent
  networks}.
\newblock \emph{\bibinfo{journal}{Nature}} \textbf{\bibinfo{volume}{464}},
  \bibinfo{pages}{1025--1028} (\bibinfo{year}{2010}).

\bibitem{CozzoPRE2012}
\bibinfo{author}{Cozzo, E.}, \bibinfo{author}{Arenas, A.} \&
  \bibinfo{author}{Moreno, Y.}
\newblock \bibinfo{title}{Stability of boolean multilevel networks}.
\newblock \emph{\bibinfo{journal}{Phys. Rev. E}} \textbf{\bibinfo{volume}{86}},
  \bibinfo{pages}{036115} (\bibinfo{year}{2012}).
\newblock \urlprefix\url{http://link.aps.org/doi/10.1103/PhysRevE.86.036115}.

\bibitem{CozzoPRE2013}
\bibinfo{author}{Cozzo, E.}, \bibinfo{author}{Ba\~nos, R.~A.},
  \bibinfo{author}{Meloni, S.} \& \bibinfo{author}{Moreno, Y.}
\newblock \bibinfo{title}{Contact-based social contagion in multiplex
  networks}.
\newblock \emph{\bibinfo{journal}{Phys. Rev. E}} \textbf{\bibinfo{volume}{88}},
  \bibinfo{pages}{050801} (\bibinfo{year}{2013}).
\newblock \urlprefix\url{http://link.aps.org/doi/10.1103/PhysRevE.88.050801}.

\bibitem{Gallotti2014}
\bibinfo{author}{Gallotti, R.} \& \bibinfo{author}{Barthelemy, M.}
\newblock \bibinfo{title}{Anatomy and efficiency of urban multimodal mobility}.
\newblock \emph{\bibinfo{journal}{Scientific reports}}
  \textbf{\bibinfo{volume}{4}} (\bibinfo{year}{2014}).

\bibitem{DeDomenico2014}
\bibinfo{author}{De~Domenico, M.}, \bibinfo{author}{Sol{\'e}-Ribalta, A.},
  \bibinfo{author}{G{\'o}mez, S.} \& \bibinfo{author}{Arenas, A.}
\newblock \bibinfo{title}{Navigability of interconnected networks under random
  failures}.
\newblock \emph{\bibinfo{journal}{Proceedings of the National Academy of
  Sciences}} \textbf{\bibinfo{volume}{111}}, \bibinfo{pages}{8351--8356}
  (\bibinfo{year}{2014}).

\bibitem{Gallotti2016}
\bibinfo{author}{Gallotti, R.}, \bibinfo{author}{Porter, M.~A.} \&
  \bibinfo{author}{Barthelemy, M.}
\newblock \bibinfo{title}{Lost in transportation: Information measures and
  cognitive limits in multilayer navigation}.
\newblock \emph{\bibinfo{journal}{Science Advances}}
  \textbf{\bibinfo{volume}{2}} (\bibinfo{year}{2016}).
\newblock \urlprefix\url{http://advances.sciencemag.org/content/2/2/e1500445}.

\bibitem{Zhang2014}
\bibinfo{author}{Zhang, L.}, \bibinfo{author}{Yang, H.}, \bibinfo{author}{Wu,
  D.} \& \bibinfo{author}{Wang, D.}
\newblock \bibinfo{title}{Solving a discrete multimodal transportation network
  design problem}.
\newblock \emph{\bibinfo{journal}{Transportation Research Part C: Emerging
  Technologies}} \textbf{\bibinfo{volume}{49}}, \bibinfo{pages}{73--86}
  (\bibinfo{year}{2014}).

\bibitem{Pang2015}
\bibinfo{author}{Pang, J. Z.~F.}, \bibinfo{author}{Bin~Othman, N.},
  \bibinfo{author}{Ng, K.~M.} \& \bibinfo{author}{Monterola, C.}
\newblock \bibinfo{title}{Efficiency and robustness of different bus network
  designs}.
\newblock \emph{\bibinfo{journal}{International Journal of Modern Physics C}}
  \textbf{\bibinfo{volume}{26}}, \bibinfo{pages}{1550024}
  (\bibinfo{year}{2015}).

\bibitem{Lin2013}
\bibinfo{author}{Lin, J.} \& \bibinfo{author}{Ban, Y.}
\newblock \bibinfo{title}{Complex network topology of transportation systems}.
\newblock \emph{\bibinfo{journal}{Transport Reviews}}
  \textbf{\bibinfo{volume}{33}}, \bibinfo{pages}{658--685}
  (\bibinfo{year}{2013}).

\bibitem{Ibarra-Rojas2015}
\bibinfo{author}{Ibarra-Rojas, O.}, \bibinfo{author}{Delgado, F.},
  \bibinfo{author}{Giesen, R.} \& \bibinfo{author}{Mu{\~n}oz, J.}
\newblock \bibinfo{title}{Planning, operation, and control of bus transport
  systems: A literature review}.
\newblock \emph{\bibinfo{journal}{Transportation Research Part B:
  Methodological}} \textbf{\bibinfo{volume}{77}}, \bibinfo{pages}{38--75}
  (\bibinfo{year}{2015}).

\bibitem{Strano2015}
\bibinfo{author}{Strano, E.}, \bibinfo{author}{Shai, S.},
  \bibinfo{author}{Dobson, S.} \& \bibinfo{author}{Barthelemy, M.}
\newblock \bibinfo{title}{Multiplex networks in metropolitan areas: generic
  features and local effects}.
\newblock \emph{\bibinfo{journal}{Journal of The Royal Society Interface}}
  \textbf{\bibinfo{volume}{12}}, \bibinfo{pages}{20150651}
  (\bibinfo{year}{2015}).

\bibitem{Gallego2010}
\bibinfo{author}{Gallego, F.~J.}
\newblock \bibinfo{title}{A population density grid of the european union}.
\newblock \emph{\bibinfo{journal}{Population and Environment}}
  \textbf{\bibinfo{volume}{31}}, \bibinfo{pages}{460--473}
  (\bibinfo{year}{2010}).

\bibitem{GoogleAPI}
\bibinfo{title}{developers.google.com/maps (accesed: 1st june 2015)}.

\bibitem{Battiston2014}
\bibinfo{author}{Battiston, F.}, \bibinfo{author}{Nicosia, V.} \&
  \bibinfo{author}{Latora, V.}
\newblock \bibinfo{title}{Structural measures for multiplex networks}.
\newblock \emph{\bibinfo{journal}{Physical Review E}}
  \textbf{\bibinfo{volume}{89}}, \bibinfo{pages}{032804}
  (\bibinfo{year}{2014}).

\bibitem{Nicosia2014}
\bibinfo{author}{Nicosia, V.} \& \bibinfo{author}{Latora, V.}
\newblock \bibinfo{title}{Measuring and modelling correlations in multiplex
  networks}.
\newblock \emph{\bibinfo{journal}{arXiv preprint arXiv:1403.1546}}
  (\bibinfo{year}{2014}).

\bibitem{Nicosia2013}
\bibinfo{author}{Nicosia, V.}, \bibinfo{author}{Bianconi, G.},
  \bibinfo{author}{Latora, V.} \& \bibinfo{author}{Barthelemy, M.}
\newblock \bibinfo{title}{Growing multiplex networks}.
\newblock \emph{\bibinfo{journal}{Physical review letters}}
  \textbf{\bibinfo{volume}{111}}, \bibinfo{pages}{058701}
  (\bibinfo{year}{2013}).

\bibitem{velocidad}
\bibinfo{note}{Average speed of each line (in spanish). Available at:
  \url{http://www.heraldo.es/noticias/zaragoza/el_tranvia_superara_menos_la_velocidad_media_los_autobuses_urbanos.html}.
  (Accesed: 1st June 2015)}.

\bibitem{frecuencia}
\bibinfo{note}{Bus frequencies in the city of Zaragoza (in spanish). Available
  at: \url{http://www.urbanosdezaragoza.es/}. (Accesed: 1st June 2015)}.

\bibitem{survey}
\bibinfo{note}{Survey: impact of tramway on mobility (in spanish). Available
  at:
  \url{http://www.ainmer.es/_files/archivos/8477b10b-96a9-458a-87f6-2c8f51b5e387.pdf}.
  (Accesed: 1st June 2015)}.

\bibitem{Distancia}
\bibinfo{note}{Guide on the public transport of Zaragoza (in spanish).
  Available at: \url{http://www.observatoriomovilidad.es/
  images/stories/08_noticias/noticia_20150521_Guia_Transporte_Publico_ZGZ.pdf}.
  (Accesed: 1st June 2015)}.

\bibitem{Browning2006}
\bibinfo{author}{Browning, R.~C.}, \bibinfo{author}{Baker, E.~A.},
  \bibinfo{author}{Herron, J.~A.} \& \bibinfo{author}{Kram, R.}
\newblock \bibinfo{title}{Effects of obesity and sex on the energetic cost and
  preferred speed of walking}.
\newblock \emph{\bibinfo{journal}{Journal of Applied Physiology}}
  \textbf{\bibinfo{volume}{100}}, \bibinfo{pages}{390--398}
  (\bibinfo{year}{2006}).

\bibitem{Mishalani2006}
\bibinfo{author}{Mishalani, R.~G.}, \bibinfo{author}{McCord, M.~M.} \&
  \bibinfo{author}{Wirtz, J.}
\newblock \bibinfo{title}{Passenger wait time perceptions at bus stops:
  Empirical results and impact on evaluating real-time bus arrival
  information}.
\newblock \emph{\bibinfo{journal}{Journal of Public Transportation}}
  \textbf{\bibinfo{volume}{9}}, \bibinfo{pages}{5} (\bibinfo{year}{2006}).

\bibitem{WatkinsFerrisBorningEtAl2011}
\bibinfo{author}{Watkins, K.~E.}, \bibinfo{author}{Ferris, B.},
  \bibinfo{author}{Borning, A.}, \bibinfo{author}{Rutherford, G.~S.} \&
  \bibinfo{author}{Layton, D.}
\newblock \bibinfo{title}{Where is my bus? impact of mobile real-time
  information on the perceived and actual wait time of transit riders}.
\newblock \emph{\bibinfo{journal}{Transportation Research Part A: Policy and
  Practice}} \textbf{\bibinfo{volume}{45}}, \bibinfo{pages}{839--848}
  (\bibinfo{year}{2011}).

\bibitem{Lagune-Reutler2015}
\bibinfo{author}{Lagune-Reutler, M.}, \bibinfo{author}{Guthrie, A.},
  \bibinfo{author}{Fan, Y.} \& \bibinfo{author}{Levinson, D.}
\newblock \bibinfo{title}{Transit riders' perception of waiting time and stops'
  surrounding enviroments}.
\newblock \emph{\bibinfo{journal}{TRANSIT}} \textbf{\bibinfo{volume}{2}},
  \bibinfo{pages}{3} (\bibinfo{year}{2015}).

\bibitem{linea2}
\bibinfo{note}{Proposal of a new line for the tramway of Zaragoza (in spanish).
  Avaiable at:
  \url{http://www.zaragoza.es/contenidos/movilidad/tranvia-linea2/general/memoria/memoria.pdf}.
  (Accesed: 1st June 2015)}.

\bibitem{Barcelona}
\bibinfo{note}{Barcelona
  \url{http://opendata.bcn.cat/opendata/es/catalog/TRANSPORT},
  \url{http://www.tmb.cat/es/linies-de-bus}. (Accesed: 1st June 2015)}.

\bibitem{Berlin}
\bibinfo{note}{Berlin
  \url{http://transitfeeds.com/p/verkehrsverbund-berlin-brandenburg/213}.
  (Accesed: 1st June 2015)}.

\bibitem{Bilbao}
\bibinfo{note}{Bilbao
  \url{ftp://ftp.geo.euskadi.net/cartografia/Transporte/Moveuskadi/}. (Accesed:
  1st June 2015)}.

\bibitem{London}
\bibinfo{note}{London \url{https://api-portal.tfl.gov.uk/docs}. (Accesed: 1st
  June 2015)}.

\bibitem{Madrid}
\bibinfo{note}{Madrid \url{http://datos.madrid.es/portal/site/egob},
  \url{http://opendata.emtmadrid.es/}. (Accesed: 1st June 2015)}.

\bibitem{Malaga}
\bibinfo{note}{Malaga \url{http://www.emtmalaga.es},
  \url{http://metromalaga.es}. (Accesed: 1st June 2015)}.

\bibitem{Valencia}
\bibinfo{note}{Valencia
  \url{http://www.valencia.es/ayuntamiento/datosabiertos.nsf},
  \url{http://www.metrovalencia.es/}, \url{https://www.emtvalencia.es/}.
  (Accesed: 1st June 2015)}.

\bibitem{Vitoria}
\bibinfo{note}{Vitoria
  \url{ftp://ftp.geo.euskadi.net/cartografia/Transporte/Moveuskadi/}.(Accesed:
  1st June 2015)}.

\bibitem{Zaragoza}
\bibinfo{note}{Zaragoza
  \url{http://www.zaragoza.es/docs-api/\#!/transporte-urbano}. (Accesed: 1st
  June 2015)}.

\bibitem{OSM}
\bibinfo{title}{www.openstreetmap.org/copyright (accesed: 1st june 2015)}.

\end{thebibliography}

\section*{Acknowledgements} A. A was supported by an FPI doctoral fellowship. This work has been partially supported by a grant to the group FENOL, by MINECO through grant FIS2014-55867-P and by the EC FET-Proactive Project Multiplex (grant 317532).

\section*{Author contributions statement}
A. A, S. M., and Y. M. designed the study, A. A. performed the analysis and the numerical simulations, A. A., S. M, and Y. M. analyzed the results. A. A, S. M. and Y. M wrote the manuscript. All authors have reviewed and approved the final version.

\textbf{Competing financial interests} The authors declare no competing financial interests.

\newpage

\begin{table}[h]
\begin{tabular}{@{}ccccccc@{}}
\toprule
\multirow{3}{*}{{\bf City}} & \multicolumn{4}{c}{{\bf Layers}}                   & \multirow{3}{*}{{\bf Nodes}} & \multirow{3}{*}{{\bf Links}} \\ \cmidrule(lr){2-5}
                              & {\bf Bus} & {\bf Metro} & {\bf Tram} & {\bf Tot.} &                              &                                \\ \cmidrule(r){1-7} 
London                        & 541        & 11         & 3          & 555        & 19459                        & 43614 \\
Berlin                     & 151       & 22          & 10         & 183        & 2866                         & 9079 \\
Madrid                        & 177       & 16          & 3          & 196        & 4703                         & 9176                       \\
Barcelona                     & 97        & 11          & 6          & 114        & 2512                         & 5337                        \\
Valencia                      & 46        & 6           & 3          & 55         & 1228  & 2617                    \\
Zaragoza                      & 35        & 0           & 1          & 36         & 915                          & 1550                        \\
Malaga                        & 41        & 2           & 0          & 43         & 1034                          & 1876                        \\
Bilbao                        & 35        & 2           & 1          & 38         & 555                          & 1212                       \\
Vitoria                       & 14        & 0           & 2          & 16         & 302                          & 452   \\ \bottomrule
\end{tabular}
\caption{ Principal characteristics of the networks under study, ordered by decreasing population\cite{Barcelona,Berlin,Bilbao,London,Madrid,Malaga,Valencia,Vitoria,Zaragoza}.}
\label{Tabla}
\end{table}

\end{document}